\documentclass[fleqn,usenatbib]{mnras}

\usepackage{newtxtext,newtxmath}

\usepackage[T1]{fontenc}

\DeclareRobustCommand{\VAN}[3]{#2}
\let\VANthebibliography\thebibliography
\def\thebibliography{\DeclareRobustCommand{\VAN}[3]{##3}\VANthebibliography}

\usepackage[normalem]{ulem}
\usepackage{graphicx}	
\usepackage{amsmath}	

\title[Stellar flares as H recombination continuum]{Hydrogen recombination continuum as the radiative model for stellar optical flares}

\author[P. J. A. Sim\~oes et al.]{
Paulo J. A. Sim\~oes$^{1,2}$\thanks{E-mail: paulo@craam.mackenzie.br (PJAS)}, 
Alexandre Araújo$^{1}$, 
Adriana Valio$^{1}$, 
Lyndsay Fletcher$^{2,4}$
\\
$^{1}$Centro de R\'adio Astronomia e Astrof\'isica Mackenzie, Escola de Engenharia, Universidade Presbiteriana Mackenzie, S\~ao Paulo, Brazil.\\
$^{2}$SUPA School of Physics and Astronomy, University of Glasgow, Glasgow G12 8QQ, UK\\
$^{3}$Rosseland Centre for Solar Physics, University of Oslo, PO Box 1029 Blindern, NO-0315 Oslo, Norway\\
}

\date{Accepted XXX. Received YYY; in original form ZZZ}

\pubyear{2023}

\begin{document}
\label{firstpage}
\pagerange{\pageref{firstpage}--\pageref{lastpage}}
\maketitle

\begin{abstract}

The study of stellar flares has increased with new observations from CoRoT, Kepler, and TESS satellites, revealing the broadband visible emission from these events. Typically, stellar flares have been modelled as $10^4$~K blackbody plasma to obtain estimates of their total energy. In the Sun, white light flares (WLFs) are much fainter than their stellar counterparts, and normally can only be detected via spatially resolved observations. Identifying the radiation mechanism for the formation of the visible spectrum from solar and stellar flares is crucial to understand the energy transfer processes during these events, but spectral data for WLFs are relatively rare, and insufficient to remove the ambiguity of their origin: photospheric blackbody radiation and/or Paschen continuum from hydrogen recombination in the chromosphere. We employed an analytical solution for the recombination continuum of hydrogen instead of the typically assumed $10^4$~K blackbody spectrum to study the energy of stellar flares and infer their fractional area coverage. We investigated 37 events from Kepler-411 and 5 events from Kepler-396, using both radiation mechanisms. We find that estimates for the total flare energy from the H recombination spectrum are about an order of magnitude lower than the values obtained from the blackbody radiation. Given the known energy transfer processes in flares, we argue that the former is a physically more plausible model than the latter to explain the origin of the broadband optical emission from flares. 
\end{abstract}

\begin{keywords}
stars: flare -- stars: solar-type -- radiation mechanisms: thermal
\end{keywords}



\section{Introduction}

Stellar magnetic activity manifests itself in a wide range of different phenomena. On the Sun, flares are high-energy events observed in the solar atmosphere. Solar flares are observed across the entire electromagnetic spectrum, such as radio, visible, ultraviolet, X-ray, and gamma rays. These transient phenomena occur in the solar atmosphere in regions of high magnetic field concentrations, where abundant amounts of energy are released in the corona by reconnection of the magnetic field \citep{benz+17,fletcher2011SSRv..159...19F}. 

Solar flares are believed to be the result of the conversion of magnetic energy into particle kinetic energy. The energy released in flares is about 10$^{27}$--10$^{32}$ erg, a large fraction of which is in the kinetic energy of accelerated, non-thermal electrons and ions, as estimated from hard X-ray and gamma-ray emission \citep[e.g.][]{Emslie2012ApJ...759...71E,Warmuth2020A&A...644A.172W}. Part of the released energy is radiated as thermal emission in soft X-rays from the corona, UV line and continuum emission from the chromosphere and from the transition region (chromosphere/corona), and white light (optical continuum) observed from the chromosphere or photosphere \citep[see e.g.][]{Milligan2014ApJ...793...70M}. Radio and millimeter emissions are also commonly detected during solar flares \citep{Bastian1998ARA&A..36..131B,white2011SSRv..159..225W}, encouraging the development of instruments to cover the sub-millimetric range, such as the Solar Submillimeter Telescope \citep[SST, ][]{Kaufmann2004ApJ...603L.121K,Kaufmann2008SPIE.7012E..0LK}, and adapting the Atacama Large Millimeter Array (ALMA) for solar observations \citep{Wedemeyer2016SSRv..200....1W,Bastian2022FrASS...9.7368B,Skokic2023A&A...669A.156S}. Therefore, observations with high temporal resolution images at all wavelengths are crucial to understanding the processes and mechanisms that occur in these complex \citep{benz2010physical}.
Solar white-light flares (WLFs), characterized by the enhancement of the optical continuum, have been extensively investigated, but often with poor spectral coverage, see e.g. \citet{Matthewsvan-Driel-GesztelyiHudson:2003} and \citet{HudsonWolfsonMetcalf:2006}. Early spectral observations, summarized by \citet{Hudson2010MmSAI..81..637H}, indicated the presence or absence of the Balmer jump, indicating the presence or absence of a hydrogen free-bound (recombination) continuum, which led the community to label WLFs as type I and type II, respectively. A lack of conclusive observations meant that several models were proposed to explain the origin of the optical continuum and the formation of solar WLF \citep{BoyerSotirovskyMachado:1985,PolandMilkeyThompson:1988,MachadoEmslieAvrett:1989}, including semi-empirical models based on observations \citep{machado1980ApJ...242..336M,mauas1990ApJ...360..715M}. In such works, one of the main points of discussion was related to the dominant mechanism producing the WL emission: a photospheric blackbody spectrum (BB) or a chromospheric H 
free-bound continuum (Hfb). However, \cite{Hudson:1972} had already pointed out the difficulty of delivering the necessary energy, by accelerated electrons, into the photosphere to heat the local plasma and form a sufficiently hot blackbody spectrum to explain the WLF observations. An alternative is to heat the photosphere via illumination of UV lines formed in the chromosphere (backwarming) during flares \citep{machado1980ApJ...242..336M}, which also suffers from severe limitations \citep{PolandMilkeyThompson:1988,simoes2017A&A...605A.125S}.

In more recent spectral observations of solar WLFs, by \citet{KerrFletcher:2014}, \citet{HeinzelKleint:2014} and \citet{KowalskiCauzziFletcher:2015}, once again both BB and Hfb models were compared, but without a definitive conclusion due to the poor spectral coverage or difficulties in obtaining an absolute calibration agreement between different instruments. Other methods were attempted to find the dominant mechanism forming the WLF emission. In the few cases where it was possible, investigations of the height of the white light emission in flares, with respect to the height of the hard X-ray (HXR) emission, were met with ambiguous results, placing the WL emission either at photospheric or mid-chromospheric heights \citep{Martinez-OliverosHudsonHurford:2012, KruckerSaint-HilaireHudson:2015}. 

While solar-dedicated instruments with large spectral coverage and resolution are not currently available, Sun-as-a-star spectra are regularly obtained by the Low-Cost Solar Telescope (LCST) coupled with the High Accuracy Radial velocity Planet Searcher for the Northern hemisphere (HARPS-N). The solar spectra are used as a reference for physical processes that drive intrinsic stellar radial-velocity variations, which interfere with the search for exoplanets using radial-velocity methods \citep{Collier2019MNRAS.487.1082C,Milbourne2019ApJ...874..107M,Pietrow2023arXiv230903373P}. However, such instruments are not typically capable of detecting continuum enhancements during solar flares, given the low contrast of the emission, which is worsened by the integration of the emission of the full solar disk. 

Dynamic models based on radiative-hydrodynamic simulations (RHD) have shown that ionization and recombination of hydrogen in the chromosphere are key factors governing the evolution of the flaring atmosphere  \citep{AbbettHawley:1999,KowalskiHawleyCarlsson:2015,simoes2017A&A...605A.125S}. In particular, \citet{simoes2017A&A...605A.125S} have shown that the dynamic process of ionization and recombination in the chromosphere, during the energy deposition phase, is fundamental in enhancing the local electron density and producing the mid-infrared (mid-IR) emission via free-free radiation \citep{ohki1975SoPh...43..405O,heinzel2012SoPh..277...31H,kaspa2009CEAB...33..309K,trottet2015SoPh..290.2809T}. Solar flare observations in the mid-IR range are becoming more common \citep{kaufmann2013ApJ...768..134K,penn2016ApJ...819L..30P,guigue2018SpWea..16.1261G,Lopes2022A&A...657A..51L,HATS2020SoPh..295...56G} and should help to place important constraints, both observational and theoretical, to identify the formation mechanism of WLFs in the Sun. 

Flares are also a common energetic phenomenon in solar-type stars and M-dwarf stars. Stellar flares have been observed in radio \citep[e.g. ][]{Bastian1987Natur.326..678B,Kundu1988A&A...195..159K} and millimetric wavelengths \citep{MacGregor2018ApJ...855L...2M,MacGregor2020ApJ...891...80M}, optical and ultraviolet ranges \citep[e.g.][]{Hawley2007PASP..119...67H}, and X-rays \citep[e.g.][]{Gudel2009A&ARv..17..309G}. More recently, multi-wavelength observations of stellar flares have become more common \citep[e.g.][]{Howard2022ApJ...938..103H,MacGregor2021ApJ...911L..25M,Namekata2023arXiv231107380N}. The interpretation of the observations were often based on knowledge and models derived from solar flare analysis \citep[e.g.][]{Hawley2003ApJ...597..535H,Maehara2015EP&S...67...59M}, including RHD simulations \citep{allred2015ApJ...809..104A,KowalskiHawleyCarlsson:2015}.
  
Kepler Space Telescope's high-precision photometry \citep{borucki2010kepler} has enabled the systematic study of stellar flares, improving our understanding of stellar activity. With the emergence of such photometric data, it was possible to have a greater temporal coverage of observations of stars \citep[e.~g.][]{daveport2016ApJ...829...23D}. For example, \citet{MaeharaShibayamaNotsu:2012} identified 365 superflares with energy on the order of 10$^{33}$ to 10$^{36}$ erg, based on 120 days of Kepler observations in 2009. These events were found in 148 G-type stars, where 14 events occurred in Sun-like stars. The results indicated that stars with superflares have an almost periodic variation in brightness, indicating the presence of very large starspots, and also that superflares in these Sun-like stars occur once every 800–5000 years \citep{shiba+13}. Following Kepler, the Transiting Exoplanet Survey Satellite \citep[TESS,][]{TESS2014SPIE.9143E..20R} has also contributed to the discovery and analysis of many more flares in active stars   \citep[e.g.][]{Feinstein2020AJ....160..219F}. An important contribution to studies of superflares was the estimation of the energy of such bursts. Many studies work with the assumption that the flare emission is originated from a blackbody plasma with $T=10^4$~K, following the conclusions of \citet{kretzschmar2011sun}. 

In this paper, we propose an analytical form for the recombination continuum of hydrogen instead of the typically assumed $10^4$~K blackbody spectrum to study the energy of stellar flares. The next section describes the data of two stars observed by the Kepler space mission. In Section~\ref{sec:models}, we detail the two proposed models for the stellar flares, whereas the results are compared in Section~\ref{sec:results}. Finally, the conclusions are presented in Section~\ref{sec:conclusions}.

\section{Data} \label{sec:style}

In this study, we use data from two stars: Kepler-411 and Kepler-396. The stellar parameters of both stars are listed in Table 1. The light curves of the two stars, Kepler-411 and Kepler-396, were retrieved from the MAST\footnote{\url{https://archive.stsci.edu/kepler/data_search/search.php}} data archive. We used short ($\approx$1 min) cadence data in the Pre-search Data Conditioning (PDCSAP) format for our analysis. There are 5 quarters of short cadence data for Kepler-411 (Q11 to Q17), and 6 quarters of Kepler-396 data (Q12 to Q17). The PDCSAP examines the calibrated light curves produced by photometric analysis and applies a series of corrections (including discontinuities, systematic trends, and outliers, such as cosmic rays) that obscure the astrophysical signals in the light curves. These corrections are based on known instrumental and spacecraft anomalies as well as unanticipated artifacts found in the data \citep{stumpe2012kepler}.

Kepler-411, a K2V star, was observed by the Kepler space telescope for about 600 days, exhibiting characteristics that indicate relatively strong magnetic activity \citep{sun2019kepler}. The activity of Kepler-411 was investigated in detail by \citet{Araujo2021ApJ...907L...5A} and \citet{AraujoFlares2021ApJ...922L..23A}. 

Kepler-396, a G star, was observed by the Kepler Space Telescope for approximately 670 days. In the analysis of Kepler-396 light curves, we identified 5 flares, which characteristics as described in Table~\ref{table396}. 

The identification of superflares in the Kepler-411 and Kepler-396 light curves was made using visual inspection of each quarter. Before performing a visual inspection of the light curves, we followed a few steps. Throughout the light curve, we checked and removed false data such as pointing errors, cosmic rays, and outliers. To remove the oscillatory trend due to the rotation of the spotted star from the light curve, a polynomial of degree three was applied and subtracted, and the relative flux of each flare with respect to the average flux of the star is obtained using Equation 2 of \citet{hawlay+14}. Then, each quarter of the light curve was visually inspected and the superflares candidates were identified as three or more consecutive points with flux above the overall average flux by at least 2.5 standard deviations $\sigma$.

\begin{figure}
\centering
  \includegraphics[scale=0.4]{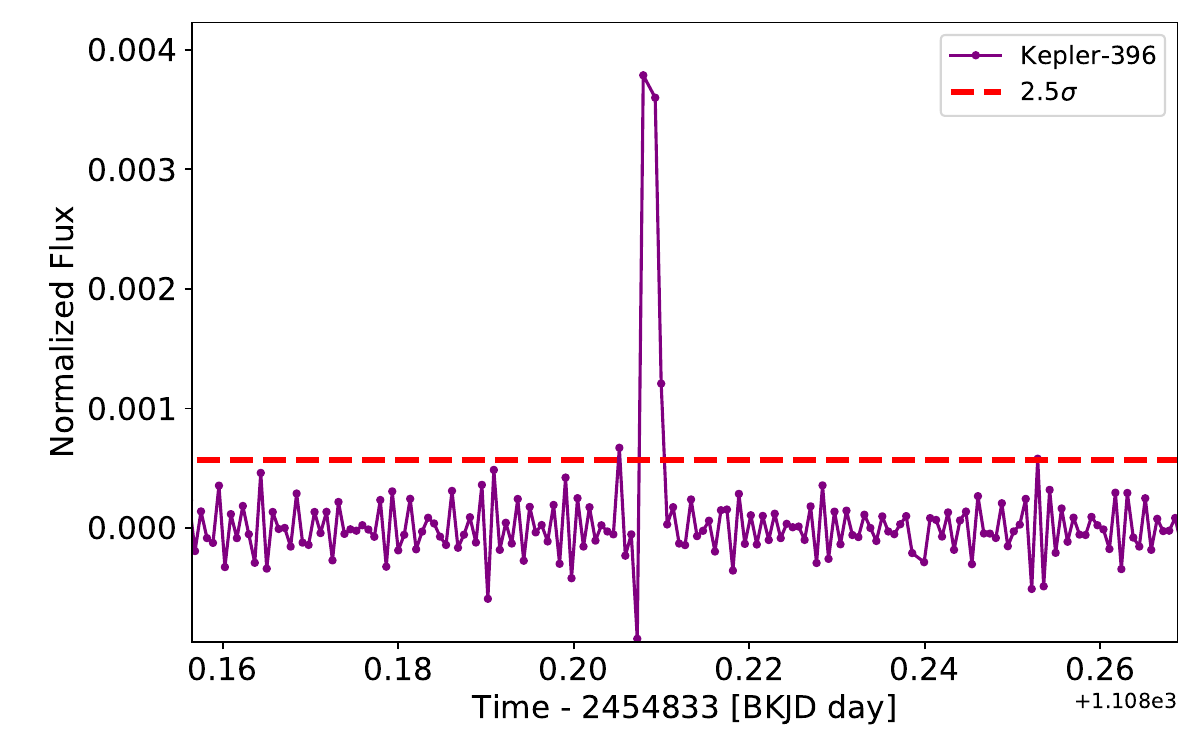}
  \caption{One of the superflares in Kepler-396 analyzed in this work, the first in Table~\ref{table396}. The lightcurve has been corrected for analysis and identification of flares and superflares (see Section~\ref{sec:style}). A sequence of three or more points above the $2.5\sigma$ threshold (red line) is considered a flare, as described in \citet{Araujo2021ApJ...907L...5A}.
    \label{fig:flare}}
\end{figure}

We analyzed 37 superflares on Kepler-411 previously analyzed by \citet{AraujoFlares2021ApJ...922L..23A} and 5 superflares on Kepler-396, using the flare models described in Section~\ref{sec:models}. To identify the superflares of the star Kepler-396 (see Figure~\ref{fig:flare}), we applied the same methodology as that used for Kepler-411, which was proposed by \citet{AraujoFlares2021ApJ...922L..23A}. 

\begin{table}
\centering
\small
\caption{Stellar parameters of Kepler-411 and Kepler-396.} \label{stars}
\begin{tabular}{l|cc}
\hline
&Stellar parameters\\

\hline
Parameter           & Kepler-411  & Kepler-396  \\
\hline
Spectral Type       & K2V$^{a}$               &  G$^{c}$ \\
Radius $R_*$ [$R_{\odot}$]& $0.820 \pm 0.018^{a}$   & 0.903$^{+0.038}_{-0.036}$$^{c}$\\
Mass [$M_{\odot}$]  & $0.87 \pm 0.04^{a}$     & 0.81$\pm$ 1.81$^{d}$\\
Effective temperature $T_\mathrm{eff}$ [K]       & 4832$^{b}$              & 5656 $\pm$ 113$^{c}$\\
Period [days]      & $10.4 \pm 0.03^{a}$      & 13.4 $^{e}$ \\
\hline
\end{tabular}
\vspace{1ex}

     {\raggedright $^{a}$\citep{sun2019kepler}; 
$^{b}$\citep{gaia2018gaia}; 
$^{c}$\citep{berger2018revised}; 
$^{d}$ \citep{xie2014transit};
$^{e}$ \citep{maehara+15}. 
\par}
\end{table}

\section{Models for the optical flare emission}
\label{sec:models} 

\subsection{Blackbody radiation}

The blackbody specific intensity $B_\lambda$ at wavelengths $\lambda$ is
\begin{equation}
    B_\lambda(T)=\frac{2hc^2}{\lambda^5} \left[ \exp 
    \left(\frac{hc}{\lambda k_B T} \right)-1 \right]^{-1}
\end{equation}
where $h$ is the Planck constant, $c$ is the speed of light, $k_B$ is the Boltzmann constant, $T$ is the temperature of the blackbody. For the quiescent stellar emission, $T$ is defined as the effective temperature of the star $T_\mathrm{eff}$. For the flare emission under the blackbody assumption, we set $T=10^4$~K, following the usual assumption made in the recent literature \cite[e.g.][]{MaeharaShibayamaNotsu:2012}. We also modelled our sample of flares using a blackbody with $T=6642$~K, which gives the same total energy output as the adopted H recombination spectrum (see Section~\ref{sec:Hfb}). 

Under these assumptions, the emitting area $A_\mathrm{f}(t)$ of an observed event can be estimated by
\begin{equation}
    A_\mathrm{f}(t)=C_\mathrm{f}(t) \pi R_*^2 \frac{\int R_\lambda B_\lambda(T_*) d\lambda}{\int R_\lambda B_\lambda(T_\mathrm{f}) d\lambda}
    \label{eq:area}
\end{equation}
where $R_*$ is the radius of the star, $R_\lambda$ is the spectral response of the telescope\footnote{The Kepler Space Telescope response function \citep{2016ksci.rept....1V} is available at \url{https://keplerscience.arc.nasa.gov/the-kepler-space-telescope.html}} and $C_\mathrm{f}(t)$ is the relative luminosity of the flare (i.e. the observed quantity obtained from Kepler data):
\begin{equation}
    C_\mathrm{f}(t)=\frac{L_\mathrm{f}(t)}{L_*}.
\end{equation}
It then follows that the flare luminosity $L_\mathrm{f}(t)$ can be estimated by using the Stefan-Boltzmann law
\begin{equation}
    L_\mathrm{f}(t) = \sigma T^4_\mathrm{f} A_\mathrm{f}(t)
\end{equation}
where $\sigma$ is the Stefan-Boltzmann constant, and $A_\mathrm{f}(t)$ is the flare emitting area, and the total radiated energy can be found by
\begin{equation}
    E_\mathrm{f} = \int_{t_0}^{t_f} L_\mathrm{f}(t) dt
    \label{eq:energy}
\end{equation}
where the time integral is done for the duration of the flare.

\subsection{Hydrogen free-bound radiation}
\label{sec:Hfb}
Following \citet{KerrFletcher:2014} and references therein, we assume a simple optically thin slab of plasma with a physical thickness $L$, with isothermal temperature $T_c$ and uniform electron density $n_e$. This slab is located above the photosphere and no radiation backwarming is considered. Under these assumptions, the hydrogen free-bound specific intensity (in erg s$^{-1}$ cm$^{-2}$ \AA$^{-1}$ sr$^{-1}$) can be calculated by \citep{1963aass.book.....A}
\begin{equation}
    I_{\lambda} = \left( \frac{6.48 \times 10^{-14}}{4\pi \lambda^2} \right)
    \left( \frac{T_c^{-3/2}}{n^3}\right) \\ \exp{\left(\frac{1.48\times 10^5}{n^2T_c} - \frac{1.44\times 10^8}{\lambda T_c}\right)}n_e^2 L
	\label{eq:hfb}
\end{equation}
where $\lambda$ is the wavelength in \AA, $n_e$ is the electron density in cm$^{-3}$, $T_c$ is the temperature in K, $L$ is the thickness of the slab in cm, $n$ is the principal quantum number of the energy level in the hydrogen atom to which the electron recombines. For the wavelength covered by Kepler observations, we consider $n=3$ (Paschen) and $n=4$ (Brackett). In this work, we adopt $T_c=10^4$~K in line with the findings of \citet{kretzschmar2011sun}, and $n_e^2L=5\times 10^{35}$~cm$^{-5}$. For reference, \citet{KerrFletcher:2014} found $n_e^2L \approx 7\times 10^{34}$~cm$^{-5}$ for the solar flare SOL2011-02-15.
As noted by \citet{KerrFletcher:2014}, Eq.~\ref{eq:hfb} assumes ionisation equilibrium, a Maxwellian velocity distribution, a pure hydrogen plasma, and the Gaunt factor is $\approx 1$. Hence, the interpretation of the results are only valid under these assumptions. We note that \citet{Machado2018ApJ...869...63M}, analyzing solar flare observations of the Lyman continuum (using data from the \emph{Extreme ultraviolet Variability Experiment} \cite[EVE, ][]{EVE2012SoPh..275..115W}, have suggested that the flaring chromospheric plasma approaches LTE conditions, supporting our adoption of the LTE assumption. A more detailed evaluation of the equilibrium assumption could be done with RHD modelling, which can handle the dynamics of ionisation and recombination during the energy deposition phase, but this is beyond our scope in this work.

\begin{figure}
	\includegraphics[width=\columnwidth]{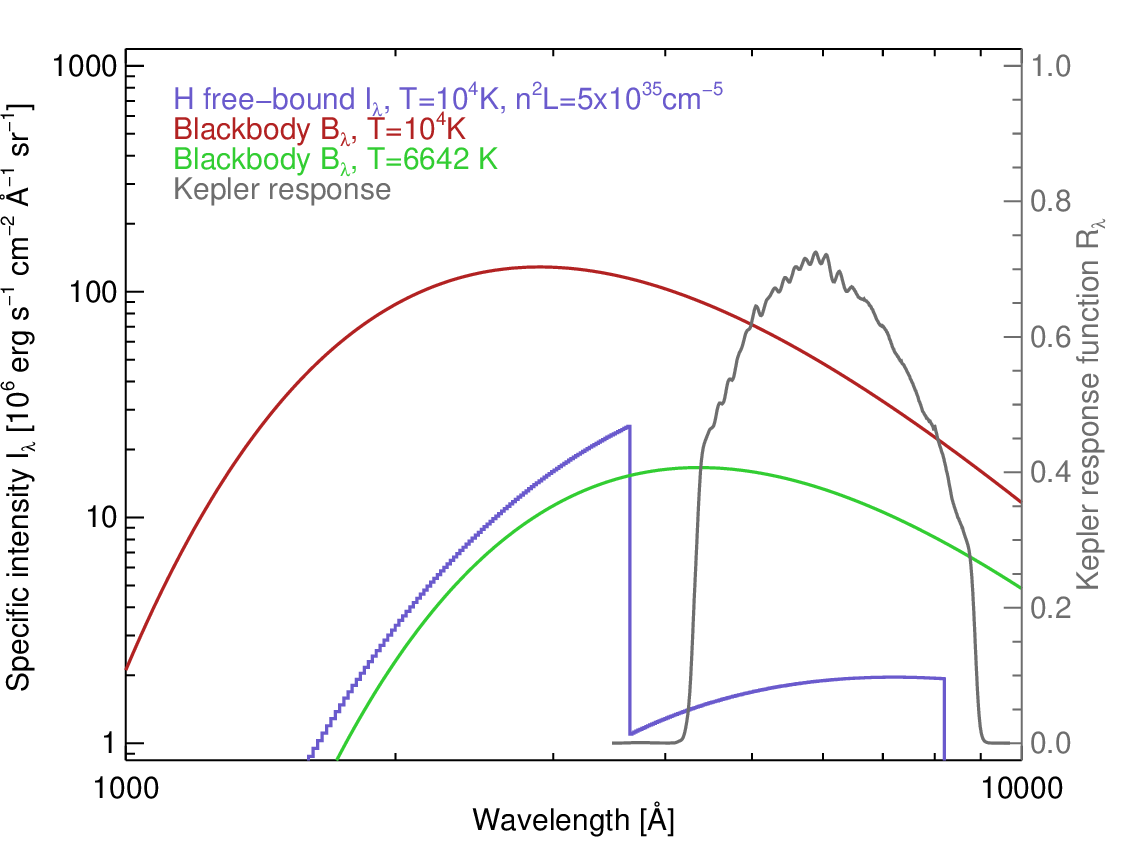}
    \caption{Flare spectral models adopted in this work: blackbody model with $T=10^4$~K (red) and $T=6642$~K (green), and the H free-bound model with $T=10^4$~K and $n^2L = 5\times 10^{35}$~cm$^{-5}$ (blue). The Kepler response function is also shown (gray).}
    \label{fig:response}
\end{figure}

Employing this radiation model to analyse the observational data is straightforward; once $I_\lambda$ is calculated, we use Eq.~\ref{eq:area}, replacing $B_\lambda(T_\mathrm{f})$ with $I_\lambda$, to obtain the flaring area $A_\mathrm{f}(t)$. The flare luminosity $L_\mathrm{f,Hfb}$ in this case is found by
\begin{equation}
    L_\mathrm{f,Hfb} = A_\mathrm{f}(t) \pi \int_{0}^{\infty} I_\lambda d\lambda, 
\end{equation}
and by performing the time integral (Eq.~\ref{eq:energy}) with $L_\mathrm{f,Hfb}$ for the duration of the event, we obtain the total radiated energy $E_\mathrm{f,Hfb}$. Note that in the following sections, we present and discuss the maximum of the relative flare area $A_\mathrm{f}(t)/(\pi R_*^2)$, i.e. the area associated with the maximum relative flux observed, for each flare. The resulting flare models are shown, along with the Kepler response function, in Figure~\ref{fig:response}.

\section{Results and discussion}\label{sec:results}

\begin{figure}
	\includegraphics[width=\columnwidth]{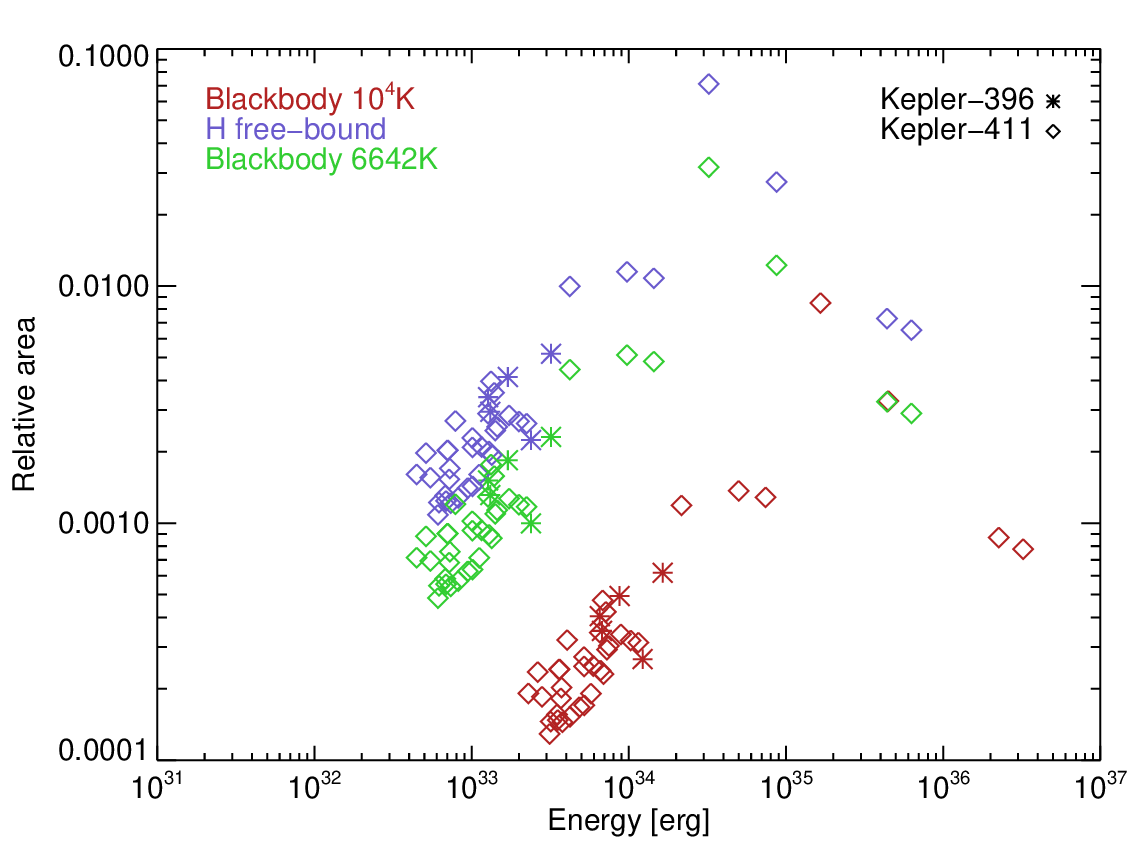}
    \caption{Relative area versus total energy   of flares from Kepler-411 (diamonds) and Kepler-396 (asterisks) stars, modelled as blackbody with $T=10^4$~K (red symbols), H free-bound continuum (blue symbols), and blackbody at $T=6642$~K (green symbols). The latter refers to a BB model that gives the same total energy as the Hfb model. Therefore, the energy estimates for the flares from the Hfb and BB at 6642~K models are the same. However, Hbf needs a larger area because Kepler would not see a large part of the Hfb spectrum (Balmer continuum), in comparison with the BB with 6642~K. }
    \label{fig:results}
\end{figure}

Our results, displayed in Figure~\ref{fig:results}, show that the Hfb model requires larger areas than the BB model (about one order of magnitude larger) and that it radiates less energy ($\approx 20\%$) than the BB at $10^4$K model. The Hfb model is a less efficient mechanism to radiate the energy away from the flaring atmosphere, compared to a BB at $10^4$K. Nevertheless, the Hfb model still yields total energy values compatible with the superflare category. The larger relative areas inferred from the Hfb model do not impose a strong constraint for the adoption of this model since spot group areas in active stars are likely to cover up at least 1\% of the surface of the star \citep{notsu2013ApJ...771..127N,okamoto2021statistical}. 

The BB model at 6642~K yields flare areas about half the size inferred from the Hfb model, for the same output energy as expected, since the temperature of 6642~K was chosen to result in the same total energy provided by the Hfb model. This exercise was performed to demonstrate that the temperature chosen for the BB model is often arbitrary, but not without an important impact on the derived parameters for the observed flares, such as the relative emitting area. The parameters for the Hfb model, namely the chromospheric temperature and emission measure $n_e^2L$, are more constrained in nature since the thickness $L$ of the emitting layer and its hydrogen density $n_e^2$ will be limited by the characteristics of the chromosphere of the star (thickness, density, and temperature), allowing for narrow possibilities for these parameters. From this, it follows that the flare areas inferred from this model vary less with the choice of the model parameters than in the case of a BB model. 

Our main argument in favor of the Hfb model instead of the BB model is based on the physics of the flaring atmosphere. The ubiquity of HXR and microwave emission in solar flares has consolidated the presence of accelerated electrons as part of the physical processes in these events \citep{fletcher2011SSRv..159...19F,white2011SSRv..159..225W}. These electron beams are commonly considered as the main process to transfer the energy released from the magnetic field into the lower atmosphere, triggering the response of the chromospheric plasma. If a similar process occurs in stellar flares, as is likely the case, the accelerated electrons cannot easily penetrate into the photosphere, instead being collisionally stopped in the chromosphere, where they deposit their energy excess \citep{Hudson:1972,HawleyFisher:1992,allred2015ApJ...809..104A}. Even if the energy can be deposited in the photosphere of the active star (by an unknown mechanism), and an originally neutral hydrogen plasma were to be heated up to $10^4$~K, the hydrogen should ionize, and if an equilibrium is maintained during the flare, the recombination continuum (Hfb) should also contribute to the flare emission. 

The backwarming mechanism, often regarded as a possible way to heat the photosphere, relies on the formation of UV lines above the photosphere. Earlier calculations by \citet{PolandMilkeyThompson:1988} suggested that backwarming illumination is not sufficient to heat the photosphere to produce WLFs. Reinforcing this conclusion, more recent RHD simulations of flares, using the RADYN code \citep{allred2015ApJ...809..104A}, do not show any changes at photospheric depths under typical flare conditions \citep[e.g.][]{simoes2017A&A...605A.125S,kerr2019ApJ...871...23K,kerr2019ApJ...883...57K,kerr2021ApJ...912..153K}.

When estimating superflare energy values, careful attention should be given to errors. This estimation of $E_\mathrm{f}$ is influenced by various types of uncertainties. Errors in stellar effective temperature ($T_\mathrm{eff}$) and stellar radius ($R$) typically affect $E_\mathrm{f}$ values by approximately 3\% and 7\%, respectively \citep{gaia2018gaia}. Determination of errors in flare start/end times and quiescent levels also impact the flare amplitude values and, consequently, $E_\mathrm{f}$ values, typically by around 30\%. We note, however, that these uncertainties affect the results of both models assumed here, and thus, do not interfere with the overall conclusions of this work.

The simple Hfb model proposed here does not take into account non-thermal ionisation of hydrogen atoms, i.e. by accelerated electrons \citep{fang1993A&A...274..917F}, which would allow for the enhancement of the free-bound continuum at temperatures below the ionisation threshold. Likewise, non-LTE effects and the dynamical evolution of the flaring atmosphere should also affect the development of the optical continuum \citep{allred2015ApJ...809..104A}.  

Another factor to consider is the variations in the flaring atmosphere during WLFs. Similarly, such variations can also pose challenges in estimating the energy of superflares, in particular under the assumption of a BB model. In the \citet{namekata2017statistical} study, the blackbody temperature of flare emissions changes to 6000-7000K, which can result in a 50\% variation in flare energy. Likewise, the chromospheric density and/or the thickness of the emitting layer is likely to vary during flares, and thus, the Hfb spectrum should vary accordingly. Furthermore, since both stellar quiescent radiation and flare emissions may not exhibit complete blackbody radiation, flare energy values (derived under a blackbody assumption) may have an error of a few tens of percent \citep{okamoto2021statistical}. Moreover, we are neglecting the presence of spectral lines in emission during flares in this analysis, but we note that they must also have some contribution to the total radiation detected by broadband photometers such as Kepler and TESS.

\cite{Maehara2015EP&S...67...59M}, analysing 187 superflares on 23 solar-type stars, found a positive correlation between flare energy and spot area (Fig. 5 of their paper), with spot areas covering up to 10\% of the stellar disk. While Maehara et al. adopted a BB flare model, a Hfb model (as presented here) would yield $\approx$ 10 times less energy, while the flare areas would still fall within the spot areas. As a consequence, the Hfb model suggests a lower requirement for the stored magnetic energy necessary to power a flare.

We also note that center-to-limb effects, i.e. variations in the observed flare intensity as a function of its location in the solar or stellar disk, may affect the emission depending on the radiation mechanism. For instance, it my have strong effects for spectral lines during flares \citep[e.g.][]{Capparelli2017ApJ...850...36C,Otsu2022ApJ...939...98O,Pietrow2023arXiv231106200PCENTRTOLIMB}. If the adopted model is a photospheric blackbody emission, then a limb-darkening effect should be accounted for, if the flare location is known - an easy task for solar flare analysis, but a difficult one for stellar flares. In the case of the chromospheric hydrogen recombination continuum model, however, such an effect might not be too relevant, although a proper radiative transfer calculations should be performed to quantify it. In any case, the parameter space of the relevant quantities (such as the electron beam parameters) should create much larger variations in the resulting spectra than the center-to-limb effects. 

\begin{table*}
\small
\centering
\caption{Parameters of Superflares in Kepler-396.} \label{table396}
\begin{tabular}{|l|l|l|l|l|l|l|}
\hline
Start  & Stop  & Duration & Rise time   & Decay time     & Peak Time     \\

[BJD]& [BJD]&  [Min]& [Min]& [Min]  & [BJD]  \\
\hline
1108.20725 & 1108.22087 & 19.6156 & 2.94192  & 16.6737   & 1108.20928   \\
 
1207.64762 & 1207.66193 & 20.5977 & 6.86592  & 13.7318   & 1207.65239  \\
    
  
1337.49850 & 1337.51552 & 24.5203 & 3.92400  & 20.5963   & 1337.50122    \\
   
  
   
1491.52902 & 1491.53992 & 15.6916 & 3.92256  & 11.7691  & 1491.53175    \\
  
   
1532.26988 & 1532.27873 & 12.7512 & 2.94192  & 9.80928   & 1532.27192   \\
\hline 
\end{tabular}
\end{table*}

\section{Conclusions}\label{sec:conclusions}

In this study, we propose the adoption of the hydrogen recombination (free-bound, Hfb) spectrum as a model for stellar flares, and present a comparison with the typical model employed for this purpose, a blackbody (BB) spectrum at $T=10^4$~K. We analyzed 37 superflares on the star Kepler-411 by \citet{AraujoFlares2021ApJ...922L..23A} and 5 superflares on the star Kepler-396, applying both models to estimate the total flare energy and relative flare area. 

We find the the Hfb model yields total flare energy estimates a factor of $\approx 10$ times smaller than the BB at $T=10^4$~K model, while requiring flare areas about 5 times larger. The inferred flare areas are still in agreement with values found by other authors and also within the estimates of spot group areas \citep{notsu2013ApJ...771..127N,okamoto2021statistical}. 

From a physical perspective, we strongly suggest the adoption of the Hfb model instead of the BB at $T=10^4$~K model. Based on decades of observations of solar flares, and models proposed to explain these observations, the bulk of the energy deposition occurs in the chromosphere. There, the predominantly neutral hydrogen can be easily ionized during flares, and it is maintained in a dynamical equilibrium, generating the Hfb continuum that should dominate over any contribution from a slightly heated photosphere - which can only marginally happen via backwarming from UV emission formed in the chromosphere. 

We strongly encourage the development of new dedicated spectrometers with high enough spectral resolution, sensitivity, and band coverage (3000 to 6000 \AA) to capture the WL emission for solar flares, and provide the much needed observational constraints for the models.

\section*{Acknowledgements}
We would like to thank the anonymous reviewer for their comments and suggestions, and Dr. Chris Osborne for discussions about LTE conditions in flares. The authors acknowledge the partial financial support received from FAPESP grants 2018/04055-8,  2021/02120-0, 2022/15700-7, CNPq grant 150817/2022-3, as well as MackPesquisa funding agency. PJAS acknowledges support from CNPq grants 307612/2019-8 and 305808/2022-2. LF acknowledges support from UK Research and Innovation's Science and Technology Facilities Council under grant award number ST/X000990/1. This manuscript
benefited from discussions held at a meeting of the International Space Science Institute team: ``Interrogating
Field-Aligned Solar Flare Models: Comparing, Contrasting and Improving'', led by Dr. G. S. Kerr and Dr. V.
Polito.
\section*{Data Availability}
The reduced data analyzed in this article is available upon request.
\bibliographystyle{mnras}
\bibliography{refs} 
\bsp	
\label{lastpage}
\end{document}